\documentclass[notitlepage,english,aps,floats,onecolumn,showpacs,nofootinbib,floatfix]{revtex4-2}

\usepackage{pslatex}
\usepackage[T1]{fontenc}
\usepackage[latin1]{inputenc}
\usepackage{graphicx}
\usepackage{epsfig}
\usepackage{longtable}
\usepackage{float}
\usepackage{calc}
\usepackage{ifthen}
\usepackage{amsmath}
\usepackage{hyperref}
\usepackage{amssymb}

\usepackage{color}

{
{
{
\newcommand{\bea}{\begin{eqnarray}}
\newcommand{\eea}{\end{eqnarray}}

\newcommand{\nc}{\newcommand}
\nc{\renc}{\renewcommand}
\nc{\eqs}[2]{\mbox{Eqs.~(\ref{#1},\,\ref{#2})}}
\nc{\eq}[1]{\mbox{Eq.~(\ref{#1})}}
\nc{\figs}[2]{\mbox{Figs.~(\ref{#1},\,\ref{#2})}}
\nc{\fig}[1]{\mbox{Fig~.(\ref{#1})}}
\nc{\be}[1]{\begin{equation} \mbox{$\label{#1}$}}
\nc{\ee}{\vspace{0.1cm}\end{equation}}

\newcommand{\bean}{\begin{eqnarray*}}
\newcommand{\eean}{\end{eqnarray*}}

\def\eV{{\rm \ eV}}
\def\GeV{{\rm \ GeV}}

\def\bfx{{\bf x}}
\def\bfk{{\bf k}}

\def\lae{\;^{<}_{\sim} \;} \def\gae{\; ^{>}_{\sim} \;}

\begin{document}

\title{Superheavy Q-Balls and Cosmology} 

\author{John McDonald }
\email{j.mcdonald@lancaster.ac.uk}
\affiliation{Dept. of Physics,  
Lancaster University, Lancaster LA1 4YB, UK}

\begin{abstract} We propose a model for the cosmological formation of superheavy Q-Balls in the mass range $10^{-7} \, M_{\odot}$ to $10^{6} \, M_{\odot}$. The model is based on a hidden sector scalar potential motivated by broken scale invariance, for which analytic Q-ball solutions and numerical simulations of condensate fragmentation exist. We show that this potential can produce superheavy Q-balls during the radiation-dominated era. As an example, we show that it is possible to produce Q-balls  of mass $ \sim \,10^{6} \, M_{\odot}$ and diameter $\sim$ 100 light years, with a number density $\sim 1$ per galaxy. Such early-forming superheavy Q-balls could play a role in galaxy and supermassive black hole (SMBH) formation. We also show that it is possible to form smaller mass Q-balls with large numbers per galaxy volume, that could form SMBH by merging. Finally, we show that it is possible to produce asteroid mass Q-balls that could account for all of the dark matter whilst remaining consistent with observational limits on MACHOs.

\end{abstract}
 
\maketitle

\section{Introduction}

 MACHOs (massive compact halo objects) are objects of mass in the range $ 10^{-7} M_{\odot}$ to $10^{6} M_{\odot}$ or more, which contribute to dark matter.  (For a summary of present observational bounds on MACHOs, see \cite{strumia}.) 
One possible class of MACHO are non-topological solitons \cite{nts}; oscillons in the case of a real scalar field $\phi$ and Q-balls in the case of a complex scalar field $\Phi$ with a conserved global charge \cite{coleman}. Their radius $R$ is dimensionally determined by the mass of the $\phi$ particle, $R \propto m_{\phi}^{-1}$, whilst their mass can be much larger than $m_{\phi}$. In particular, a Q-ball, which is absolutely stable due to global charge conservation, can carry an extremely large global charge $Q$ and have a correspondingly large mass $M_{Q} \approx Q m_{\phi}$, whilst having a radius $R_{Q} \propto m_{\phi}^{-1}$ that is independent of $Q$. 

JWST has observed black-hole powered quasars with mass $\gae 10^{10} M_{\odot}$ at surprisingly high redshift, $z \geq 6$, when the Universe was less than 950 million years old \cite{jwst}. It has also observed more galaxies at high redshift than predicted by the standard cosmological model \cite{jwst2}.   
MACHOs, if they are sufficiently massive and form sufficiently early, could modify the evolution of galaxies and the formation of supermassive black holes (SMBH).

Here we explore the formation in the early Universe of superheavy Q-balls of mass $10^{-7} M_{\odot}$ to $10^{6}  M_{\odot}$. We will show that, for a plausible scalar potential with a broken shift symmetry, it is possible to consistently form superheavy Q-balls during the radiation-dominated era. In particular, we will show that it is possible to produce Q-balls of mass $10^{6} M_{\odot}$ and diameter $\sim$ 100 light years, with a number density of the order of one per galaxy, that could act as a seed for SMBH and galaxy formation.
We also show that it is possible to form a much larger number density of lighter Q-balls, with a much larger number per galaxy volume, that could subsequently form SMBH by merging. This is possible because the Q-ball radius does not change as its mass increases, whereas its Schwarzschild radius increases, eventually forming a black hole. Finally, 
we show that it is possible to produce asteroid mass Q-balls, in the mass range $10^{-16} M_{\odot}$ to $10^{-11} M_{\odot}$, that could account for all of the dark matter whilst remaining consistent with observational limits on MACHOs.

\section{Q-balls} 

A Q-ball is the minimum energy configuration of a complex scalar field with a fixed global charge $Q$. For a complex scalar field $\Phi(\bfx, t)$ with a global $U(1)$ symmetry, the energy and charge of a field configuration are
\be{qb1} E = \int \left(|\dot{\Phi}|^2 + |\nabla\Phi|^2 + V(\Phi)\right) d^{3}x ~\ee
and
\be{qb2} Q = i \int \left( \dot{\Phi}^{\dagger} \Phi - \Phi^{\dagger} \dot{\Phi} \right) d^{3}x   ~.\ee 
A configuration of the form 
\be{qb3} \Phi(\bfx, t) = \frac{\phi(\bfx)}{\sqrt{2}} e^{i \omega t}  ~\ee
has charge 
\be{qb4} Q =  \omega \int \phi^{2}\left(\bfx\right) \, d^{3}x    ~\ee
and energy  
\be{qb5} E =  \int \left( \frac{|\nabla \phi(\bfx)|^{2}}{2}  + \frac{\omega^2 \phi^{2}(\bfx) }{2} + V(\phi) \right) \, d^{3}x    ~.\ee
The minimum energy configuration will be spherically symmetric, $\phi(\bfx) = \phi(r)$. The charge and energy are then 
 \be{qb6} Q =  4 \pi \omega \int_{0}^{\infty} \phi^{2}\left(r\right) r^2 \, dr    ~\ee
and  
\be{qb5a} E = 4 \pi \int_{0}^{\infty} \left( \frac{1}{2} \left(\frac{\partial\phi(r)}{\partial r} \right)^{2} + \frac{\omega^2 \phi^{2}(r) }{2} + V(\phi) \right) r^{2} dr    ~.\ee
The equation of motion\footnote{Here $\Phi$ and $\Phi^{\dagger}$ are treated as independent variables in $V(\Phi, \Phi^{\dagger})$.  This is valid when determining the field equation (see \cite{colemannotes}, p.56). }
\be{qb6a} \ddot{\Phi} - \nabla^{2} \Phi = - \frac{\partial V(\Phi, \Phi^{\dagger})}{\partial \Phi^{\dagger}}  ~,\ee  
becomes 
\be{qb7} \frac{\partial^{2} \phi}{\partial r^{2}} + \frac{2}{r} \frac{\partial \phi}{\partial r} = \frac{\partial V}{\partial \phi} - \omega^{2} \phi \equiv 
\frac{\partial V_{\omega}}{\partial \phi} ~\ee
where
\be{qb8} V_{\omega}(\phi) = V(\phi) - \frac{\omega^{2} \phi^{2}}{2} ~.\ee 
We refer to \eq{qb7} as the Q-ball equation. This is solved with boundary conditions $\phi'(0) = 0$  and $\phi(r) \rightarrow 0$ as $r \rightarrow \infty$, assuming that the vacuum expectation value is zero. 

To form a stable Q-ball there must be an attractive interaction between the scalar particles, so that the energy per charge of the Q-ball, $E/Q$, is less than the mass of the isolated scalar particles. It would then require additional energy to disperse the Q-ball scalars, ensuring absolute stability of the Q-ball if $\Phi$ is the lowest mass particle carrying global charge. An attractive interaction corresponds to a scalar potential that increases less rapidly than $|\Phi|^{2}$ as $|\Phi|$ increases from zero \cite{coleman}. 

\section{Perturbation Growth and Condensate Fragmentation} 

\subsection{Perturbation growth}

Q-balls can form due to growth of perturbations of an initial condensate of complex scalars. If the initial condensate is globally charged (Q-matter), then growth of the perturbations of the condensate will fragment the condensate into globally charged lumps that will relax to Q-balls. On the other hand, if the initial condensate is neutral, corresponding to a complex scalar oscillating along a line in the complex plane, then the growth of perturbations will fragment the condensate to initially neutral lumps that will relax to oscillons. Numerical simulations of this process have shown that the initial oscillons subsequently break up into a distribution of positive and negative charge Q-balls \cite{gravnum}.

In the following we will consider an initially neutral complex scalar condensate. This allows Q-balls to form with a simple potential, since there is no need to generate an initial asymmetry in the condensate.

A complete understanding of condensate fragmentation requires numerical simulation. However, for some potentials, and in particular the potential we will be considering, it is possible to use perturbation theory to determine the size of the fragments that form and the time of their formation. 

  The analytical method was originally discussed in \cite{lee} for the case of flat space and subsequently applied to the case of supersymmetric (SUSY) flat direction scalars in the expanding Universe with gauge-mediated SUSY breaking in \cite{bs}, and with gravity-mediated SUSY breaking in \cite{qb2}. The potential we will consider here is identical in form to the flat direction potential of gravity-mediated SUSY breaking at small $|\Phi|$ \cite{qb2,qb1}. 

The analysis in \cite{bs} is specifically for the case of perturbations of Q-matter, which is a circularly rotating complex scalar condensate that corresponds to a maximally charged condensate, essentially a condensate made of purely $Q = 1$ or $Q = -1$ scalars. However, the physics of condensate fragmentation is more general than the nature of the scalars. It can be physically understood as the formation of a spherical configuration of non-relativistic scalars due to the attractive interaction between the scalars. Therefore we expect a similar perturbation growth rate for the case of real scalars as for the case of complex scalars with the same potential. In the following we will review the analysis of \cite{bs} to determine the growth rate of perturbations of a Q-matter condensate and the size of the resulting condensate fragments. We will then apply the resulting perturbation growth rate to the case of both real and complex condensates.

The complex field is\footnote{In \cite{bs} the field is written as $\Phi =  R e^{i \Omega}$, and the potential is written as $U(R)$, which is equivalent to $V(\Phi)/2$. We will convert the analysis of \cite{bs} to the more conventional notation in terms of $\Phi$ and $V(\Phi)$.} 
\be{qb9}  \Phi(\bfx,t) = \frac{1}{\sqrt{2}} \phi(\bfx, t)e^{i \Omega(\bfx, t)}  ~.\ee 
The complex field equation is equivalent to two real equations
\be{qb10} \ddot{\phi} + 3 H\dot{\phi} - \frac{\nabla^{2} \phi}{a^{2}} - \phi \dot{\Omega}^{2} + \phi \frac{\left(\nabla \Omega\right)^{2}}{a^{2}} = -\frac{\partial V}{\partial \phi}   ~\ee 
and 
\be{qb11} \ddot{\Omega} + 3 H \dot{\Omega} + \frac{2 \dot{\phi} \dot{\Omega}}{\phi} - \frac{2 (\nabla \phi).(\nabla \Omega)}{\phi a^{2}}  - \frac{\nabla^{2} \Omega}{a^{2}}  = 0   ~.\ee 
We next consider perturbations around a homogeneous background, with $\phi(\bfx, t) = \phi(t) + \delta \phi(\bfx, t)$ and $\Omega(\bfx, t) = \Omega(t) + \delta \Omega(\bfx, t) $. The homogeneous equations are 
\be{qb12a} \ddot{\phi} +3 H \dot{\phi} - \phi \dot{\Omega}^{2} = - \frac{\partial V}{\partial \phi} ~\ee
and 
\be{qb13a} \ddot{\Omega} + \left( 3 H + \frac{2 \dot{\phi}}{\phi} \right)  \dot{\Omega} = 0 ~,\ee
and perturbation equations are 
\be{qb12} \delta \ddot{\phi} + 3 H \delta \dot{\phi} - \frac{\nabla^{2}\delta \phi}{a^{2}}  - 2 \phi \dot{\Omega} \delta \dot{\Omega} - \dot{\Omega}^{2} \delta \phi = - V''(\phi) \, \delta \phi   ~\ee
and   
\be{qb13} \delta \ddot{\Omega} + 3 H \delta \dot{\Omega} +  2 \frac{\dot{\phi}}{\phi} \delta \dot{\Omega} +  2 \frac{\delta \dot{\phi}}{\phi} \dot{\Omega}  - \frac{2 \dot{\phi} \dot{\Omega}}{\phi^{2}} \delta \phi - \frac{\nabla^{2} \delta \Omega}{a^{2}}  = 0 ~.\ee
The solution is assumed to be of the form $\delta \phi$, $\delta \Omega \propto e^{S(t)} e^{-i \bfk.\bfx}$. In addition, it is assumed that $\dot{S} = \alpha$, a constant, where $\alpha$ is the perturbation growth rate. Then \eq{qb12} and \eq{qb13} become 
\be{qb14} \delta \phi \left[ \alpha^{2} + 3 H \alpha + 
\frac{k^2}{a^2} - \dot{\Omega}^{2} + V'' \right] + \delta \Omega \left[-2 \phi \dot{\Omega} \alpha \right] = 0  ~\ee
and 
\be{qb15} \delta \phi \left[ \frac{2 \dot{\Omega} \alpha}{\phi} - \frac{2 \dot{\phi} \dot{\Omega}}{\phi^{2}} \right] + 
\delta \Omega \left[\alpha^{2} + 3 H \alpha + 2 \frac{\dot{\phi}}{\phi} \alpha + \frac{k^{2}}{a^{2}}  \right]  = 0 ~.\ee
These equations can then be combined into a dispersion relation 
\be{qb16} \left[\alpha^{2} + 3 H \alpha + \frac{k^{2}}{a^{2}} + \frac{2 \dot{\phi}}{\phi} \alpha \right]\left[\alpha^{2} + 3 H \alpha + \frac{k^{2}}{a^{2}} -\dot{\Omega}^{2} + V'' \right]
+ 4 \dot{\Omega}^{2} \alpha^{2} \frac{\dot{\phi}}{\phi}    = 0 ~.\ee 
To solve this, $\alpha$ is assumed to be large compared to $H$ and $\dot{\phi}/\phi$, in which case we can neglect expansion when calculating $\alpha$ at a given time. (For Q-matter, $\phi$ changes only due to expansion, which dilutes the charge density. Therefore $\dot{\phi}/\phi \approx H$ and it is sufficient that $\alpha$ is large compared to $H$.) Then \eq{qb16} becomes
\be{qb18} \alpha^{4} + \alpha^{2} \left(3 \dot{\Omega}^{2} +  \frac{2 k^{2}}{a^{2}} + \frac{V''}{2} \right) + \frac{k^{2}}{a^{2}} \left(\frac{k^{2}}{a^{2}} - \dot{\Omega}^{2} + V'' \right) = 0 ~.\ee 
For a solution to exist, it is necessary that
\be{qb19} \frac{k^{2}}{a^{2}} \, < \, \frac{k_{max}^{2}}{a^{2}}   ~\ee
where  
\be{qb19a} \frac{k_{max}^{2}}{a^{2}} = \dot{\Omega}^{2} - V''   ~.\ee 
In the limit where expansion can be neglected, we have $\dot{\phi} = H = 0$ and so, from \eq{qb12a},  
\be{qb19b} \dot{\Omega}^{2} = \frac{V'}{\phi} ~.\ee
Therefore 
\be{qb19c} \frac{k_{max}^{2}}{a^{2}} =  \frac{V'}{\phi} - V''   ~.\ee 
\eq{qb18} can be written in terms of $k_{max}/a$ as 
\be{qb20} \alpha^{4} + \alpha^{2} \left( 4 \dot{\Omega}^{2} +  \frac{2 k^{2}}{a^{2}} - \frac{k_{max}^{2}}{a^2} \right) + \frac{k^{2}}{a^{2}}\left(\frac{k^{2}}{a^{2}} - \frac{k_{max}^{2}}{a^{2}} \right) = 0 ~,\ee
with solution
\be{qb21} \alpha^{2} = \frac{1}{2} \left[- \left(4 \dot{\Omega}^{2} + \frac{2 k^{2}}{a^{2}} - \frac{k_{max}^{2}}{a^{2}}\right) + \sqrt{ 8 \dot{\Omega}^{2}\left(2 \dot{\Omega}^{2} + \frac{2 k^{2}}{a^{2}} - \frac{k_{max}^{2}}{a^{2}} \right) + \frac{k_{max}^{4}}{a^{4}} }  \right] ~.\ee  
$\alpha$ goes to zero at $k = 0$ and at $k = k_{max}$. 
For $4 \dot{\Omega}^{2} \gg k_{max}^{2}/a^{2}$, which is true in the cases of interest, the solution to leading order becomes 
\be{qb22} \alpha = \frac{ \frac{k}{a} \left( \frac{k_{max}^{2}}{a^{2}} - \frac{k^{2}}{a^{2}} \right)^{1/2}}{2 \dot{\Omega}}  ~.\ee 
The growth factor $S$ for a given mode $k$ is then 
\be{qb23} S = \int_{t_{k,\,i}}^{t} \alpha dt   ~\ee
where $t_{k,\,i}$ is the time at which $k^{2}/a^{2}$ becomes smaller than $k_{max}^{2}/a^{2}$, with corresponding scale factor $a_{k,\,i}$. 

We will show in the next section that for the potential of interest to us here, which is dominated by the mass squared term, $\dot{\Omega}$ is a constant. In this case $S$ becomes
\be{qb23a} S = \int_{t_{k,\,i}}^{t} \frac{ \frac{k}{a} \left( \frac{k_{max}^{2}}{a^{2}} - \frac{k^{2}}{a^{2}} \right)^{1/2}}{2 \dot{\Omega}} \,dt = \frac{1}{2 \dot{\Omega}} \int_{a_{k,\,i}}^{a} \frac{k}{a} \left( \frac{k_{max}^{2}}{a^{2}}  -  \frac{k^{2}}{a^{2}}\right)^{1/2} \frac{da}{aH}   ~.\ee
We will assume that fragmentation occurs during radiation domination, in which case $H \propto 1/a^{2}$. Then 
\be{qb24} S = \frac{1}{2 \dot{\Omega} H}  \left( \frac{k_{max}}{a} \right)^{2} F(a) ~,\ee
where 
\be{qb25} F(a) = \left(\frac{a_{k,\,i}}{a}\right)  \left[ \sqrt{1- \left(\frac{a_{k,\,i}}{a}\right)^{2} } - \left(\frac{a_{k,\,i}}{a}\right)  \tan^{-1}\left(\sqrt{\frac{1 - \left(\frac{a_{k,\,i}}{a}\right)^{2}}{\left(\frac{a_{k,\,i}}{a}\right)^{2} } } \right)  \right]   ~.\ee
We find that $F(a)$ is maximised at $a_{k,\,i}/a = 0.39 \approx 2/5$ (where the latter will be a convenient approximation for analytical expressions). The corresponding maximum value of $F(a)$ is  $F(a) = 0.18 \approx 1/5$.  
This gives the value of $a_{k,\,i}$ and $F(a)$ for the mode that has the most growth at a given $a$. The maximum growth factor at a given $a$, which we denote by $S_{*}$, is 
\be{qb26} S_{*} = \frac{1}{10 \dot{\Omega} H} \left( \frac{k_{max}}{a}\right)^{2}   ~.\ee 
Since $k/a = k_{max}/a$ at $a_{k,\,i}$, the perturbation mode with the  maximum growth at a given $a$, $k_{*}$, is
\be{qb27} \frac{k_{*}}{a} = \left(\frac{a_{k,\,i}}{a}\right) \left(\frac{k_{max}}{a}\right) = \frac{2}{5} \left(\frac{k_{max}}{a}\right) ~.\ee
This will give the wavenumber of the mode that fragments the condensate. The corresponding physical wavelength, which will give the diameter of the condensate fragments, is 
\be{qb28} \lambda_{*} = \frac{2 \pi}{k_{*}/a} = \frac{5 \pi}{k_{max}/a}   ~.\ee 

There are two possible sources for the initial scalar field perturbation: conventional adiabatic perturbations from inflaton quantum fluctuations and isocurvature perturbations due to $\phi$ quantum fluctuations. For now we will assume that the adiabatic perturbations are dominant. For perturbation modes that enter the horizon after the field begins oscillating (at $ H < H_{i} \approx m_{\phi}/3 $) but for which $k/a > k_{max}/a$ at a given $a$ and so there is no growth due to the attractive interaction, the magnitude of the perturbation will not change from its primordial value since the sub-horizon perturbations have $k/a < m$ and are therefore non-relativistic, in which case $\phi$ and $\delta \phi$ have the same $a^{-3/2}$ dependence. Since $\rho_{\phi} \propto \phi^{2}$, it follows that 
\be{qb29} \left(\frac{\delta \rho_{\phi}}{\rho_{\phi}}\right)_{0} = 2 \left(\frac{\delta \phi}{\phi}\right)_{0} \approx 5 \times 10^{-4} ~, \ee 
where we consider the primordial adiabatic density perturbations to have magnitude $5 \times 10^{-4}$. As $a$ increases, eventually $k/a$ becomes smaller that $k_{max}/a$ and the perturbation begins to grow due to the attractive interaction. This begins at $a_{k,\,i}$. Then  
\be{qb30}  \frac{\delta \phi}{\phi} = \left(\frac{\delta \phi}{\phi} \right)_{0} e^{S}   ~.\ee
The condensate will fragment once $\delta \phi/\phi \approx 1$, which occurs once when $S$ satisfies 
\be{qb31}  S = -\ln \left(\frac{\delta \phi}{\phi}\right)_{0} \equiv \gamma_{frag} \approx 10.6   ~.\ee 
This will be satisfied first for the mode that has the maximum value of $S$ at a given $a$ , $S_{*}$. Therefore, using \eq{qb26}, we find for the value of $H$ at condensate fragmentation, $H_{frag}$,  
\be{qb33} H_{frag} = \frac{\left(\frac{k_{max}}{a}\right)^{2} }{10 \dot{\Omega} \gamma_{frag} } ~.\ee

 The preceding analysis is for the growth of perturbations of   
a Q-matter condensate made of scalars with $Q = 1$ or $Q = -1$. The advantage of the Q-matter condensate is that the time dependence of the background condensate can be factored out, allowing an analytical solution for the growth of condensate perturbations. However, since the physics of perturbation growth of a complex rotating Q-matter condensate and a real coherently oscillating condensate is essentially the same, we expect essentially the same growth of the perturbations\footnote{The potential of interest to us here, which is discussed in the next section, is dominated by a $|\Phi|^{2}$ term, in which case the rotating condensate is, to a good approximation,  equivalent to two real coherently oscillating condensates in the $\phi_{1}$ and $\phi_{2}$ direction which are $\pi/2$ out of phase with each other, where $\Phi = (\phi_{1} + i \phi_{2})/\sqrt{2}$. The only difference compared to a real condensate is an additional attractive interaction between the $\phi_{1}$ and $\phi_{2}$ scalars, in addition to the attractive self-interactions of the  $\phi_{1}$ and $\phi_{2}$ scalars.}. This has been confirmed in \cite{jn1,jn2,jn3}, where the Q-matter analysis is adapted to the growth of perturbations of a real oscillating condensate by averaging over the oscillations of the condensate. Therefore we will also apply the perturbation growth rate of the Q-matter analysis to the case of interest to us here, in which the initial condensate has no global charge.

\subsection{Non-linear evolution} 

In \cite{gravnum}, the growth of perturbations of a scalar condensate that is neutral or close to neutral, corresponding to field trajectories in the complex plane that are highly elliptical and close to linear oscillations, was studied numerically for the gravity-mediated type potential. (This case had previously been studied numerically in \cite{kk}.) 
The results were that the almost neutral condensate first fragmented into almost neutral objects  (called first generation Q-balls in \cite{gravnum}), which correspond to oscillons in the case of a completely neutral initial condensate. These oscillons then evolve into a distribution of positive and negative Q-balls. 

The significance of this is that it is not necessary for a global charge asymmetry to exist in the initial condensate in order to form Q-balls. This will allow a relatively simple and plausibly natural potential to generate Q-balls in the early Universe, as we discuss in the next section.

\section{A model for the cosmological formation of superheavy Q-balls}

We consider a complex scalar $\Phi$ in a hidden sector with a global $U(1)$ symmetry which does not directly interact with Standard Model particles. The potential consists of a mass term plus a logarithmic correction, 
\be{e1} V(\Phi) = m_{\phi}^{2}|\Phi|^{2} - K m_{\phi}^2 |\Phi|^{2} \ln\left(\frac{2 |\Phi|^{2}}{\mu^{2}} \right)  ~,\ee 
where $K > 0$. $\mu$ is the renormalisation scale in the case where the logarithmic term arises via quantum corrections. More generally, even when $V(\Phi)$ is considered purely classically, a change in $\mu^2$ simply rescales $m_{\phi}^2$ and $K$ whilst keeping the same form for $V(\Phi)$, so $\mu$ can be considered arbitrary when $m_{\phi}^2$ and $K$ are free parameters, in which case  $m_{\phi}^2$ and $K$ are defined relative to a given choice of $\mu$. This form of this potential is the same as that of a MSSM flat direction in the case of gravity-mediated SUSY breaking \cite{qb2,qb1,asko}. As reviewed below, the great advantage of this potential is that analytic solutions for the Q-balls exist, as first shown in \cite{qb2}.  We note that this form of potential is natural if there is a shift symmetry of the potential, $\Phi \rightarrow \Phi + C$, where $C$ is an arbitrary complex constant, with the shift symmetry-breaking parameter being $m_{\phi}^{2}$. In this case all terms in the potential, including quantum corrections, must be proportional to $m_{\phi}^2$. This also explains the absence of a $\lambda |\Phi|^4$ term. 

\section{Analytical Q-ball solution} 

The Q-ball equation for this potential is 
\be{e2} \frac{\partial^{2} \phi}{\partial r^2} + \frac{2}{r} \frac{\partial \phi}{\partial r} = V'(\phi) - \omega^{2} \phi = m_{\phi}^{2}\left(1 - K  - K\ln\left(\frac{\phi^{2}}{\mu^{2}}\right)\right) \phi   -\omega^{2} \phi ~.\ee 
This can be solved by substituting
\be{e4} \phi(r) = \phi_{0} \exp \left(- B r^2\right) ~,\ee
which gives 
\be{e4a} 4 B^{2} r^{2} - 6 B = 2 B K m_{\phi}^2 r^2 +  m_{\phi}^2 - \omega^2 - K m_{\phi}^2 \left(1 + \ln\left(\frac{\phi_{0}^2}{\mu^2}\right) \right) ~.\ee
Therefore
\be{e5} B = \frac{K m_{\phi}^{2}}{2} ~\ee
and 
\be{e6}  B = \frac{1}{6} \left( \omega^{2} - m_{\phi}^{2} + K m_{\phi}^{2} \left(1 + \ln\left(\frac{\phi_{0}^{2}}{\mu^{2}}\right) \right) \right) ~.\ee
\eq{e5} and \eq{e6} imply that 
\be{e7} \omega^{2} = m_{\phi}^{2} + K m_{\phi}^{2}\left(2 - \ln\left( \frac{\phi_{0}^{2}}{\mu^{2}} \right) \right) ~.\ee  
We define $K$ and $m_{\phi}$ with respect to the choice $\mu = \phi_{0}$, in which case 
\be{e8} \omega^{2} = m_{\phi}^{2} (1 + 2 K) ~.\ee
 Therefore
\be{e6a} \phi(r) = \phi_{0}\, \exp\left(-\frac{K m_{\phi}^{2} r^{2}}{2}\right) ~\ee
and the complete Q-ball solutions is   
\be{e6b} \Phi(\bfx, t) = \frac{\phi_{0}\,\exp \left(i \omega t\right) }{\sqrt{2}}  \exp\left(-\frac{K m_{\phi}^{2} r^{2}}{2}\right)    ~.\ee
The charge and mass of the Q-ball are then
\be{e9} Q = 4 \pi \omega \phi_{0}^{2} \int_{0}^{\infty} r^{2} e^{-K m_{\phi}^{2} r^{2}} \, dr = 
\frac{\pi^{3/2} \phi_{0}^{2} \omega}{K^{3/2} m_{\phi}^{3}}   =  \frac{\pi^{3/2} \phi_{0}^{2} \left(1 + 2 K\right)^{1/2}}{K^{3/2} m_{\phi}^{2}}   ~\ee
and
\be{e10} M_{Q} \equiv E = 4 \pi \phi_{0}^{2} \int_{o}^{\infty} r^{2} \left( \frac{m_{\phi}^{2} + \omega^{2}}{2} + K^{2} m_{\phi}^{4} r^{2} \right) e^{-K m_{\phi}^{2} r^{2}} \, dr
= \frac{ \pi^{3/2} \phi_{0}^{2} \left(1 + \frac{5 K}{2} \right)}{ K^{3/2} m_{\phi} }   ~.\ee 
We define the radius of the Q-ball, $R_{Q}$, as the distance at which $\phi(R_{Q}) = \phi_{0}/e$,
\be{e11} R_{Q} = \frac{\sqrt{2}}{\sqrt{K} m_{\phi}}    ~.\ee 

The above analysis assumes that the attractive self-interaction of the Q-ball scalars dominates over gravity. In the Appendix we show that a conservative condition for this to be true is that $\phi_{0}/M_{Pl} < K$.

\section{Q-ball Cosmology} 

\subsection{Fragmentation and Q-ball formation}

As discussed in Section 3, we will calculate the perturbation growth for a Q-matter condensate with potential \eq{e2} and then apply this to the case of the initially neutral condensate.  

The field will become dynamical and start evolving once $ H \approx m_{\phi}/3$, at which time the expansion rate is $H_{i}$ and the scale factor is $a_{i}$. For the potential of \eq{e1}, $\dot{\Omega}^{2}$ is given by 
\be{f1} \dot{\Omega}^{2} = \frac{V'}{\phi} = m_{\phi}^{2} \left(1 - K - K \ln \left(\frac{\phi^{2}}{\mu^{2}} \right) \right) = 
m_{\phi}^{2} \left(1 - K\right) ~,\ee
where we have set $\mu = \phi$ in the last expression. 
$k_{max}^{2}/a^{2}$ is, from \eq{qb19c}, 
\be{f1a} \frac{k_{max}^{2}}{a^{2}} = 2 K m_{\phi}^{2}  ~.\ee
Then, from \eq{qb33}, the condensate will fragment once the expansion rate equals 
\be{f2} H_{frag} = \frac{K m_{\phi}}{5 (1 - K)^{1/2} \gamma_{frag}}   ~.\ee  
From \eq{qb28} the diameter of the condensate lumps when the condensate fragments is
\be{f3} \lambda_{frag} = \lambda_{*} = \frac{5 \pi}{\sqrt{2 K} m_{\phi}}  ~.\ee
The scale factor when the condensate fragments, $a_{frag}$, is
\be{f4} \frac{a_{frag}}{a_{i}} = \left(\frac{H_{i}}{H_{frag}}\right)^{1/2} = \left(\frac{5 (1 - K)^{1/2} \gamma_{frag}}{3 K} \right)^{1/2}    ~,\ee
where we have used $H_{i} = m_{\phi}/3$ and radiation domination is assumed. 
The field at fragmentation is then related to the initial value of the field when the field starts evolving,  $\phi_{i}$, by 
\be{f5} \phi_{frag} = \left(\frac{a_{i}}{a_{frag}}\right)^{3/2} \phi_{i} = \left(\frac{H_{frag}}{H_{i}}\right)^{3/4} \phi_{i} = \left(\frac{3 K}{5 \left(1 - K \right)^{1/2} \gamma_{frag}}\right)^{3/4} \phi_{i}  ~.\ee
The fragmenting perturbation is well within the horizon for any $K$ that is small compared to 1,  
\be{f5x} \lambda_{frag} < H^{-1}_{frag} \Rightarrow \gamma_{frag} \approx 10.6 > \frac{\pi \sqrt{K}}{(2 (1-K))^{1/2}} ~.\ee
It is also straightforward to confirm that $\alpha \gg H$ at fragmentation, as assumed in our derivation of the growth rate. From \eq{qb22}, we find that at fragmentation, 
\be{f5x2} \alpha \approx \frac{0.35 \left(\frac{k_{max}}{a}\right)^{2}}{\dot{\Omega}} \approx \frac{0.7 K m_{\phi}}{(1 - K)^{1/2}}  \gg H_{frag} \Leftrightarrow \gamma_{frag} \gg 0.3 ~.\ee

Initially, the fragments will form into oscillons, as the coherently oscillating condensate has no global charge. Once the oscillons form, it is known from numerical simulation of the potential \eq{e1} that the oscillons are unstable and will evolve into positive and negative Q-balls \cite{gravnum}. The results of \cite{gravnum} were that an almost neutral condensate (ratio of minor to major axis $\epsilon = 0.01$) first fragments into almost neutral objects  (called first generation Q-balls in \cite{gravnum}) which are essentially oscillons. These oscillons then evolve into a distribution of positive and negative Q-balls. For the case $\epsilon = 0.01$ and $K = 0.1$, the distribution is dominated by large $\pm$ Q-ball pairs with charge $Q \approx 3 \times 10^{-4} \phi_{i}^{2}/m_{\phi}^{2}$ (where we set $|\Phi_{in}|^{2}$ in \cite{gravnum} to $\phi_{i}^{2}/2$ in our notation) at the peak of the distribution (see Fig.25 of \cite{gravnum}). 
This is smaller than the charge that would be expected if the initial oscillons fragmented into a single pair of $\pm$ Q-balls with $\phi_{0} \approx \phi_{frag}$. If we compute the charge of the Q-balls assuming that $\phi_{0} = \phi_{frag}$, then using \eq{e9} we obtain 
\be{f6a} Q = \frac{\pi^{3/2} (1 + 2K)^{1/2}}{(1 - K)^{3/4} \gamma_{frag}^{3/2}} \left(\frac{3}{5}\right)^{3/2} \frac{\phi_{i}^{2}}{m_{\phi}^{2}}   ~.\ee 
With $\gamma_{frag} = 16.1$ (consistent with the initial value for perturbations $\delta \phi/\phi = 10^{-7}$ used in \cite{gravnum}) and $K = 0.1$, we find $Q \approx 0.04 \phi_{i}^{2}/m_{\phi}^{2}$, which is about 130 times larger than the dominant Q-balls from the numerical simulation. In our analysis we will simplify the initial distribution of Q-balls to an ensemble of $\pm$ Q-ball pairs with mass and charge that are smaller than the value $E$ and $Q$ calculated assuming that $\phi_{0} = \phi_{frag}$ by a factor $\eta_{Q}$. Therefore we will use 
\be{f6} \phi_{0} = \eta_{Q}^{1/2} \phi_{frag}  ~,\ee
with $\eta_{Q}$ in the range 0.001 to 0.1, when discussing the properties of the resulting Q-balls.

\subsection{Present Q-ball Density}

In the following we will assume that the comoving energy in the initial oscillating scalar field and  in the final state of Q-balls are, to a good approximation, the same. This is likely as the binding energy per scalar in the Q-balls, $E_{B}/Q$, given in \eq{ap4a},  is small compared to the rest mass of the scalars, with $E_{B}/Q \sim K m_{\phi}$ for $K$ small compared to 1. So we expect relatively little energy to be lost due to the reconfiguration of the condensate scalar particles into Q-balls.

Under the assumption that the energy density in the oscillating scalar field becomes an energy density in Q-balls at $a_{frag}$, which subsequently evolves as non-relativistic matter, the present density in Q-balls is simply what the present density in the oscillating scalar field would be if it did not fragment, 
\be{c16} \rho_{Q,\,0} = \left(\frac{a_{i}}{a_{0}}\right)^{3} \rho_{\phi,\,i} = \frac{1}{2} \frac{g(T_{0})}{g(T_{i})} \frac{T_{0}^{3}}{T_{i}^{3}} m_{\phi}^{2} \phi_{i}^{2}   ~\ee  
and
\be{c17a} \Omega_{Q,\,0} =  \frac{g(T_{0})}{g(T_{i})} \frac{T_{0}^{3}}{T_{i}^{3}} \frac{m_{\phi}^{2} \phi_{i}^{2}}{2 \rho_{c,\,0}}  ~,\ee  
where $\rho_{c,0}$ is the present critical density. 

The temperature at which the $\phi$ field begins oscillating is
\be{c17b} H_{i} = \frac{k_{T_{i}} T_{i}^{2}}{M_{Pl}} = \frac{m_{\phi}}{3}  \Rightarrow 
T_{i} = \frac{\left(m_{\phi} M_{Pl}\right)^{1/2}}{(3 k_{T_{i}})^{1/2}}  ~,\ee 
where $k_{T}  =  \left(\pi^{2} g(T)/90\right)^{1/2}$.  
Therefore
\be{c17c} \Omega_{Q,\,0} = \frac{g(T_{0})}{g(T_{i})}  \frac{\left(3 k_{T_{i}}\right)^{3/2}}{2} \frac{T_{0}^{3} m_{\phi}^{1/2} \phi_{i}^{2} }{M_{Pl}^{3/2} \rho_{c,\,0} }    ~.\ee
 Then from \eq{f5}, 
\be{c17d} \Omega_{Q,\,0} = \frac{g(T_{0})}{g(T_{i})} \left(\frac{k_{T_{i}}^{3/2}}{2}\right) 
\frac{T_{0}^{3} m_{\phi}^{1/2} M_{Pl}^{1/2}}{\rho_{c,\,0} } \left(\frac{5 (1 - K)^{1/2} \gamma_{frag}}{K}\right)^{3/2} 
\left(\frac{\phi_{frag}}{M_{Pl}}\right)^{2}   ~.\ee
It will be useful for our analysis to express $\Omega_{Q,\,0}$ in terms of $\phi_{0}/M_{Pl}$ 
\be{c17e} \Omega_{Q,\,0} = \frac{g(T_{0})}{g(T_{i})} \left(\frac{k_{T_{i}}^{3/2}}{2}\right) 
\frac{T_{0}^{3} m_{\phi}^{1/2} M_{Pl}^{1/2}}{\eta_{Q} \rho_{c,\,0} } \left(\frac{5 (1 - K)^{1/2} \gamma_{frag}}{K}\right)^{3/2} 
\left(\frac{\phi_{0}}{M_{Pl}}\right)^{2}   ~.\ee

\subsection{Superheavy Q-balls} 

We define $N_{\odot}$ as the mass of the superheavy Q-ball in solar masses, 
\be{c17} M_{Q} = N_{\odot} M_{\odot}   ~,\ee
where $M_{\odot} = 1.1 \times 10^{57} \GeV$.  We will express our results in terms of the present Q-ball density $\Omega_{Q,\,0}$. To simplify our expressions we will assume from now on that $K$ is small compared to 1.

Using \eq{e10}, the mass of the scalar is related to $\phi_{0}/M_{Pl}$ by 
\be{c18} m_{\phi} = \frac{\pi^{3/2} \phi_{0}^{2}}{K^{3/2} M_{Q} } = \frac{\pi^{3/2} M_{Pl}^{2}}{K^{3/2} N_{\odot} M_{\odot}  } \left(\frac{\phi_{0}}{M_{Pl}}\right)^{2}  ~,\ee 
which gives
\be{c19} m_{\phi}  = \frac{2.9 \times 10^{-20} }{K^{3/2} N_{\odot}} \left(\frac{\phi_{0}}{M_{Pl}}\right)^{2}\, \GeV   ~.\ee 
Substituting \eq{c19} into \eq{c17e} gives 
\be{c20} \Omega_{Q,\,0} = \frac{g(T_{0})}{g(T_{i})} \frac{\pi^{3/4} k_{T_{i}}^{3/2}}{2 \eta_{Q}} 
\frac{(5 \gamma_{frag})^{3/2}}{K^{9/4}} 
\frac{T_{0}^{3} M_{Pl}^{3/2}}{\rho_{c,\,0} \, N_{\odot}^{1/2}M_{\odot}^{1/2} } \left(\frac{\phi_{0}}{M_{Pl}}\right)^{3} ~.\ee 
With $g(T_{i}) = 100$, $g(T_{0})  = 3.91$, $k_{T_{i}} = 3.3$, $T_{0} = 2.4 \times 10^{-13} \GeV$ and $\rho_{c,\,0} = 4.0 \times 10^{-47} \GeV^{4}$, this becomes
\be{c21} \Omega_{Q,\,0} =  
\frac{4.1 \times 10^{9}}{\eta_{Q} K^{9/4} N_{\odot}^{1/2}} \left( \frac{\gamma_{frac}}{10.6}\right)^{3/2}  \left(\frac{\phi_{0}}{M_{Pl}}\right)^{3} ~.\ee 
Therefore 
\be{c25}  \frac{\phi_{0}}{M_{Pl}} = 6.3 \times 10^{-4} \; \Omega_{Q,\,0}^{1/3} \eta_{Q}^{1/3} \left( \frac{10.6}{\gamma_{frag} }\right)^{1/2}  K^{3/4} N_{\odot}^{1/6}  ~.\ee 
Substituting \eq{c25} into \eq{c19} then gives 
\be{c19a} m_{\phi} = 1.1 \times 10^{-26} \; \frac{\Omega_{Q,\,0}^{2/3} \eta_{Q}^{2/3}}{N_{\odot}^{2/3}} \left(\frac{10.6}{\gamma_{frag}}\right) \; \GeV ~.\ee
The radius of the Q-ball is 
\be{c22} R_{Q} = \frac{\sqrt{2}}{\sqrt{K} m_{\phi}} 
= \frac{\sqrt{2} K N_{\odot} M_{\odot}}{\pi^{3/2} M_{Pl}^{2}} 
\left(\frac{\phi_{0}}{M_{Pl}}\right)^{-2} ~,\ee 
Substituting \eq{c25}, we obtain 
\be{c27} R_{Q} = 1.3 \times 10^{26} \; \left( \frac{\gamma_{frag}}{10.6} \right)  
\frac{N_{\odot}^{2/3}}{\eta_{Q}^{2/3} \Omega_{Q,\,0}^{2/3} K^{1/2}}  \; \GeV    \equiv  2.4 \times 10^{7} \;   \left( \frac{\gamma_{frag}}{10.6} \right)  
\frac{N_{\odot}^{2/3}}{\eta_{Q}^{2/3} \Omega_{Q,\,0}^{2/3} K^{1/2}} \; {\rm km} ~.\ee
The fragmentation temperature is 
\be{c23} T_{frag} = \left(\frac{H_{frag} M_{Pl}}{k_{T_{frag}}}\right)^{1/2} ~.\ee
Using \eq{f2} for $H_{frag}$, this becomes 
\be{c23a} T_{frag} = \left(\frac{K m_{\phi} M_{Pl}}{5 \gamma_{frag} k_{T_{frag}}}\right)^{1/2} ~.\ee
Then using \eq{c19a} to eliminate $m_{\phi}$, we obtain
\be{c26} T_{frag} = 1.2 \times 10^{-5} \;  \Omega_{Q,\,0}^{1/3} \, \eta_{Q}^{1/3} \left( \frac{10.6}{\gamma_{frag}} \right)  
\frac{K^{1/2}}{N_{\odot}^{1/3}} \; \GeV ~.\ee

\subsection{Present spacing between Q-balls due to expansion} 

At fragmentation, the distance between the initial fragments (neutral oscillons) is $\lambda_{frag}$, given by \eq{f3}. A fragment will evolve into a number of positive and negative Q-balls, with the number per initial fragment being $\approx 1/\eta_{Q}$. Therefore, assuming the positive and negative Q-balls are initially equally spaced (which appears consistent with the spatial distribution observed in the numerical simulation of \cite{gravnum}), the initial separation of the Q-balls will be 
\be{c28a} \lambda_{Q,\,i} = \eta_{Q}^{1/3} \lambda_{frag} ~.\ee
The present separation of the Q-balls purely due to expansion will then be 
\be{c28} \lambda_{Q,\,0} \approx \left(\frac{a_{0}}{a_{frag}} \right) \lambda_{Q,\,i} = \frac{5 \pi \eta_{Q}^{1/3}}{\sqrt{2 K} m_{\phi}} \left(\frac{g(T_{frag})}{g(T_{0})} \right)^{1/3} \frac{T_{frag}}{T_{0}} ~.\ee
Therefore 
\be{c29} \lambda_{Q,\,0} = 1.46 \times 10^{35} \; \left(\frac{N_{\odot}}{\Omega_{Q,\,0}}\right)^{1/3} \, \GeV^{-1} \equiv \,
0.94 \; \left(\frac{N_{\odot}}{\Omega_{Q,\,0}}\right)^{1/3} \,{\rm kpc}  ~. \ee

\section{Examples of Superheavy Q-balls in cosmology} 

The present bounds on MACHOs are presented in Figure 3.3 of \cite{strumia}. For MACHO masses in the range $10^{-11} M_{\odot}$ to $10^{5} M_{\odot}$, MACHOs are generally allowed by observational limits on microlensing and accumulation in the galactic centre if the fraction of dark matter in MACHOs, $f$, is less than $10^{-3}$. This upper bound becomes relaxed towards the upper end of the $M_{\odot}$ range, where, for MACHOs with mass close to $10^{4} M_{\odot}$,  $f$ as large as $0.1$ is allowed by microlensing. An interesting window exists for asteroid mass MACHOs in the range $10^{-16} M_{\odot}$ to $10^{-11} M_{\odot}$ where MACHOs could account for all of dark matter, although there are recent bounds from GRB lensing that may increase the lower bound on this range from $10^{-16} M_{\odot}$ to $10^{-14} M_{\odot}$ \cite{strumia}. The relation between $f$ and  $\Omega_{macho}$ is  $\Omega_{macho} = \Omega_{DM} f = 0.26 f$.

\subsection{One superheavy Q-ball per galaxy with $M_{Q} \approx 10^{6} M_{\odot}$ } 

We first consider the possibility that a superheavy Q-ball of mass of the order of $10^{6} M_{\odot}$ could form with an expansion spacing similar to the length scale of galaxy-sized  perturbations, $\sim 1$ Mpc, so that there will be around one superheavy Q-ball per galaxy. Since the superheavy Q-balls will be seen to form very early (close to matter-radiation equality), we speculate that these Q-balls may act as seeds for the early formation of SMBH and galaxies.

From \eq{c29}, we find
\be{ex1}  \Omega_{Q,\,0} \approx \left(\frac{0.94\,{\rm kpc} }{\lambda_{Q,\,0}} \right)^{3} N_{\odot} \approx 8.3 \times 10^{-4} \, \left(\frac{N_{\odot}}{10^{6}} \right) \left(\frac{1 \,{\rm Mpc} }{\lambda_{Q,\,0}} \right)^{3}  ~.\ee
For $N_{\odot} = 10^{6}$ this is essentially the same as the observational upper limit on $10^{6} M_{\odot}$ MACHOs, which is $f \approx 2 \times 10^{-3} \Rightarrow \Omega_{macho} \approx 5.2 \times 10^{-4}$ (from Fig 3.3 of \cite{strumia}). 

Using \eq{c26}, the fragmentation temperature is given by 
\be{ex2} T_{frag} = 0.77 \; \left(\frac{\eta_{Q}}{0.01}\right)^{1/3} \left(\frac{1 \,{\rm Mpc} }{\lambda_{Q,\,0}} \right)
\left( \frac{10.6}{\gamma_{frag}} \right)
\left(\frac{K}{0.1} \right)^{1/2} \, \eV  ~,\ee
where we normalise our numbers to the case $K = 0.1$, which is a reasonable estimate for a dimensionless parameter. Therefore fragmentation occurs mostly during radiation domination but ends close to matter-radiation equality.

From \eq{c27}, the radius of the Q-ball is 
\be{ex3}  R_{Q} = 1.8 \times 10^{15} \; \left(\frac{\lambda_{Q,\,0}}{1 \,{\rm Mpc} } \right)^{2} 
\left( \frac{\gamma_{frag}}{10.6} \right)
\left(\frac{0.01}{\eta_{Q}}\right)^{2/3}
\left(\frac{0.1}{K} \right)^{1/2} \,{\rm km}  \equiv 196  \; \left(\frac{\lambda_{Q,\,0}}{1 \,{\rm Mpc} } \right)^{2} 
\left( \frac{\gamma_{frag}}{10.6} \right)
\left(\frac{0.01}{\eta_{Q}}\right)^{2/3}
\left(\frac{0.1}{K} \right)^{1/2} \,{\rm l.y.} ~.\ee
Thus the mass and radius of the $10^{6} M_{\odot}$ Q-ball are similar to the mass and radius of a globular cluster.

We note that the fragmentation temperature and the radius of the Q-ball are independent of $N_{\odot}$ and so independent of $M_{Q}$ if we require one Q-ball per galaxy volume ($\sim$ 1 ${\rm Mpc}^{3}$).  Therefore we could consider any mass of Q-ball and the fragmentation temperature would still be close to matter-radiation equality and the radius of the Q-ball would still be of the order of 100 l.y. .

The mass of the scalar is, from \eq{c19a}, 
\be{sm5} m_{\phi} = 4.6 \times 10^{-25} \; 
 \left(\frac{\eta_{Q}}{0.01}\right)^{2/3}
\left(\frac{10.6}{\gamma_{frag}}\right)
\left(\frac{1 \,{\rm Mpc} }{\lambda_{Q,\,0}} \right)^{2} \,{\rm eV} 
~.\ee

The Q-ball field is, from \eq{c25},  
\be{sm4} \phi_{0} = 5.6 \times 10^{13} \; 
 \left(\frac{\eta_{Q}}{0.01}\right)^{1/3}
\left(\frac{N_{\odot}}{10^{6}}\right)^{1/2} 
\left(\frac{10.6}{\gamma_{frag}}\right)^{1/2}
\left(\frac{K}{0.1}\right)^{3/4} 
\left(\frac{1 \,{\rm Mpc} }{\lambda_{Q,\,0}} \right) \, \GeV 
~.\ee 
The initial scalar field amplitude is then, from \eq{f5} and \eq{f6}, 
\be{sc15} \phi_{i} = \left( \frac{5 \gamma_{frag}}{3 K} \right)^{3/4} \frac{\phi_{0}}{\eta_{Q}^{1/2}} = 2.7 \times 10^{17} \; 
 \left(\frac{0.01}{\eta_{Q}}\right)^{2/3}
\left(\frac{\gamma_{frag}}{10.6}\right)^{1/4}
\left(\frac{N_{\odot}}{10^{6}}\right)^{1/2} 
\left(\frac{1 \,{\rm Mpc} }{\lambda_{Q,\,0}} \right) \,\GeV 
~.\ee

We have assumed that the $\phi$ perturbations are dominated by adiabatic perturbations. As with any ultra-light particle, isocurvature perturbations are expected due to the quantum fluctuations of the light scalar during inflation, $\delta \phi_{iso} \approx H_{I}/(2\pi)$ where $H_{I}$ is the expansion rate during inflation. For the adiabatic perturbation to dominate, we therefore require that 
\be{sm6} 2 \left(\frac{\delta \phi_{iso}}{\phi_{i}}\right) = \frac{H_{I}}{\pi \phi_{i}} \lae 5 \times 10^{-5} ~.\ee
Therefore, if $\phi_{i} \approx 3 \times 10^{17} \GeV$ then 
$H_{I} \lae 5 \times 10^{13} \GeV$ is necessary for the adiabatic perturbation to dominate. However, as long as the isocurvature perturbation of the Q-ball density, which contributes less than $10^{-3}$ of the dark matter density, does not violate CMB bounds on the isocurvature perturbation, a dominant Q-ball isocurvature perturbation would only slightly modify the value of $\gamma_{frag}$ and so would not significantly alter our conclusions.

\subsection{SMBH formation from Q-ball mergers} 

We next discuss the possibility that the collision and merger of Q-balls in a galaxy could lead to the formation of SMBH. 
The Schwarzschild radius of the Q-ball is 
\be{sc1} R_{s} = \frac{M_{Q}}{4 \pi M_{Pl}^{2}} = \frac{N_{\odot} M_{\odot}}{4 \pi M_{Pl}^{2}}   ~.\ee 
Therefore, with $R_{Q}$ given by \eq{e11}, we obtain 
\be{sc2} \frac{R_{s}}{R_{Q}} = \frac{\sqrt{K} N_{\odot} M_{\odot} m_{\phi}}{4 \sqrt{2} \pi M_{Pl}^{2} }   ~.\ee  
Thus, using \eq{c19a}, this becomes
\be{sc3} \frac{R_{s}}{R_{Q}} = 1.2 \times 10^{-7} \; K^{1/2} N_{\odot}^{1/3} \Omega_{Q,\,0}^{2/3} \eta_{Q}^{2/3} \left(\frac{10.6}{\gamma_{frag}}\right)   ~.\ee
A significant property of the Q-balls we are considering here is that their radius is independent of their mass. Therefore, if Q-balls merge into larger Q-balls, the mass increases but the radius remains the same. On the other hand, the Schwarzschild radius increases as the Q-ball mass increases. We can then estimate the number, $N_{merge}$, of Q-balls that would have to merge for $R_{s}/R_{Q}$ to increase to 1 and so for the resulting Q-ball to collapse into a black hole,
\be{sc4} N_{merge} = \frac{1}{R_{s}/R_{Q}} = \frac{8.2 \times 10^{6}}{ K^{1/2} N_{\odot}^{1/3} \Omega_{Q,\,0}^{2/3} \eta_{Q}^{2/3} \left(\frac{10.6}{\gamma_{frag}}\right)   }  
 ~.\ee 
It is only energetically favourable for like-sign Q-balls to merge, since the energy per unit charge will decrease as the total charge of the Q-ball increases. So, in effect, opposite charges repel but like charges attract and merge when it comes to an ensemble of positively and negatively charged Q-balls. We will therefore assume that over time Q-ball collisions only increase the charge of the Q-balls. 

The number of Q-balls in a galaxy volume is 
\be{sc5} N_{Q,\,gal} \approx \left(\frac{1 \, {\rm Mpc}}{\lambda_{Q,\,0}}\right)^{3} = \left(\frac{1 \, {\rm Mpc}}{0.94 \,{\rm kpc}}\right)^{3} \frac{\Omega_{Q,\,0}}{N_{\odot}} = 1.2 \times 10^{9} \, \frac{\Omega_{Q,\,0}}{N_{\odot}}  ~.\ee 
The total mass in Q-balls in a galaxy volume is then 
\be{sc6} M_{Q,\,gal} = N_{Q,\,gal} N_{\odot} M_{\odot} = 1.2 \times 10^{9} \, M_{\odot}  \Omega_{Q,\,0} ~.\ee
We assume that a black hole forms once a fraction $\epsilon_{merge}$ of the Q-balls in the galaxy volume have merged. Then the black hole mass is 
\be{sc7} M_{BH} = \epsilon_{merge} M_{Q,\,gal} =  1.2 \times 10^{9} \, M_{\odot}  \epsilon_{merge} \Omega_{Q,\,0} ~,\ee
where $\epsilon_{merge} \leq 1$. We will apply the conservative MACHO upper bound on the fraction of dark matter due to Q-balls, $f \leq 10^{-3}$, corresponding to $\Omega_{Q,\,0} \leq 2.6 \times 10^{-4}$. Normalising $\Omega_{Q,\,0}$ to this upper bound, we obtain  
\be{sc8} M_{BH} =  3.1 \times 10^{5} \, M_{\odot} \, \epsilon_{merge} \left(\frac{\Omega_{Q,\,0}}{2.6 \times 10^{-4}} \right)  ~.\ee
Thus the largest mass of a black hole from Q-ball mergers, corresponding to the case where all of the Q-balls within the galaxy volume merge, $\epsilon_{merge} = 1$,  is $10^{5} - 10^{6} \, M_{\odot}$.

We next determine the properties of the Q-balls and of the scalar field required to form the black hole from the merger of a fraction $\epsilon_{merge}$ of the Q-balls in the galaxy volume. The number of mergers necessary to form the black hole is  
\be{sc9} N_{merge} = \epsilon_{merge} N_{Q,\,gal}  ~.\ee 
From \eq{sc4} and \eq{sc5}, the mass of the Q-balls in solar mass units is then
\be{sc10} N_{\odot} = \frac{M_{Q}}{M_{\odot}} = 1.8 \times 10^{3} \epsilon_{merge}^{3/2} K^{3/4} \Omega_{Q,\,0}^{5/2}  \eta_{Q} \left(\frac{10.6}{\gamma_{frag}}\right)^{3/2} ~.\ee
Therefore the mass of the Q-balls is 
\be{sc11} M_{Q} = 7.0 \times 10^{21}  \; \epsilon_{merge}^{3/2} \left( \frac{K}{0.1}\right)^{3/4} \left( \frac{\Omega_{Q,\,0}}{2.6 \times 10^{-4}}\right)^{5/2} \left( \frac{\eta_{Q}}{0.01} \right) \left(\frac{10.6}{\gamma_{frag}}\right)^{3/2} \, {\rm kg}    ~.\ee
Thus the Q-ball mass for the case $\epsilon_{merge} = 1$ is similar to the mass of the Moon, $M_{Moon} = 7.4 \times 10^{22}$ kg.

The Q-ball radius is, from \eq{c27} and \eq{sc10},  
\be{sc12}  R_{Q} = 3.5 \times 10^{9} \;
 \epsilon_{merge} \Omega_{Q,\,0} \, {\rm km}   = 9.2 \times 10^{5} \; \epsilon_{merge} \left( \frac{\Omega_{Q,\,0}}{2.6 \times 10^{-4}}\right)\, {\rm km}  ~.\ee
Thus the Q-ball radius for $\epsilon_{merge} = 1$ is similar to the solar radius, $R_{\odot} \approx 7 \times 10^{6} $ km. 

The mass of the scalar, from \eq{c18} and \eq{sc10}, is
\be{sc12a} m_{\phi} = \frac{9.4 \times 10^{-25}}{\epsilon_{merge}}  \, \left( \frac{2.6 \times 10^{-4}}{\Omega_{Q,\,0}}\right)\left( \frac{0.1}{K}\right)^{1/2}\; \GeV ~.\ee

The Q-ball field is 
\be{sc14} \frac{\phi_{0}}{M_{Pl}} = 1.7 \times 10^{-3} \, \epsilon_{merge}^{1/4} \Omega_{Q,\,0}^{3/4} \eta_{Q}^{1/2} K^{7/8} \left(\frac{10.6}{\gamma_{frag}}\right)^{3/4}    ~,\ee
which gives 
\be{sc14a} \phi_{0} = 1.4 \times 10^{11} \; \epsilon_{merge}^{1/4} \left( \frac{\Omega_{Q,\,0}}{2.6 \times 10^{-4}}\right)^{3/4} \left( \frac{\eta_{Q}}{0.01} \right)^{1/2} \left( \frac{K}{0.1}\right)^{7/8} \left(\frac{10.6}{\gamma_{frag}}\right)^{3/4} \; \GeV ~.\ee
The initial scalar field amplitude is then 
\be{sc15x} \phi_{i}  = \left( \frac{5 \gamma_{frag}}{3 K} \right)^{3/4} \frac{\phi_{0}}{\eta_{Q}^{1/2}} = 7.0 \times 10^{13} \; \epsilon_{merge}^{1/4} \left(\frac{K}{0.1}\right)^{1/8}  \left( \frac{\Omega_{Q,\,0}}{2.6 \times 10^{-4}}\right)^{3/4} \, \GeV ~.\ee 
In this case, for the adiabatic perturbation initial condition to dominate over the isocurvature perturbation, we require that $H_{I} \lae 4 \times 10^{10} \GeV$ for $\epsilon_{merge} = 1$. 

As the Q-balls merge, the radius of the merged Q-ball remains the same but the field in the Q-ball increases. The Q-ball field is related to the Q-ball mass by \eq{c18}. Substituting the scalar mass \eq{sc12a} in \eq{c18}, we find that the Q-ball mass is related to the Q-ball field by 
\be{sc15a} M_{Q} = 9.8 \times 10^{5} M_{\odot} \epsilon_{merge} 
\left( \frac{0.1}{K}\right)
\left( \frac{\Omega_{Q,\,0}}{2.6 \times 10^{-4}}\right) \left( \frac{\eta_{Q}}{0.01} \right)^{1/2}  \left(\frac{\phi_{0}}{M_{Pl}}\right)^{2} ~.\ee
Therefore, when the Q-ball reaches the mass at which it collapses into a black hole, \eq{sc8}, the Q-ball field is 
\be{sc15b} \frac{\phi_{0}}{M_{Pl}} = 0.56 \left( \frac{K}{0.1}\right)^{1/2}  ~.\ee
Thus the Q-ball field is close to the Planck mass when the black hole forms. This raises two issues. Firstly, as shown in the Appendix, gravity becomes as important as the attractive scalar self-interaction once $\phi_{0}/M_{Pl} \approx K$, which is true for all $K \lae 3$ when \eq{sc15b} is satisfied. However, the effect of gravity will only enhance the merging of Q-balls and the formation of a black hole, so this will not qualitatively alter our conclusions. The other issue is that Planck-suppressed potential terms might modify the scalar potential. For example, a shift symmetry-breaking term of the form $\lambda (m_{\phi}/M_{Pl})^{2} |\Phi|^{4} $ could become comparable to the mass squared term in the potential as $\phi \rightarrow M_{Pl}$. Whether this is really a problem will depend upon the details of the complete quantum gravity theory. We will assume that any such corrections do not significantly modify our results.

Thus Q-balls of roughly lunar mass and solar radius could merge to produce a black hole of mass $10^{5}-10^{6} M_{\odot}$ assuming that all the Q-balls in the galaxy volume merge. This initial black hole could subsequently grow by accretion of conventional and dark matter. If the Q-ball mass is smaller then a smaller fraction $\epsilon_{merge}$ of the 
Q-balls need to merge to form a black hole (since $N_{merge} \propto N_{\odot}^{-1/3}$ and $N_{Q,\,gal} \propto N_{\odot}^{-1} \Rightarrow \epsilon_{merge} \propto N_{\odot}^{2/3}$). In this case the initial black holes from merger will have a smaller mass. Nevertheless, the initial black holes could subsequently grow via accretion to produce larger mass black holes.  

The above discussion assumes that the Q-balls can merge sufficiently rapidly. We will discuss the conditions for this to be true in subsection D.

\subsection{Asteroid mass Q-balls as dark matter} 

Q-balls in the mass range $10^{-14}-10^{-11} M_{\odot}$ can account for all of the dark matter whilst remaining consistent with observational limits on MACHOs \cite{strumia}. To study this case we set $\Omega_{Q,\,0} = 0.26$. Then the radius of the Q-ball, from \eq{c27}, is 
\be{as1} R_{Q} = 30.9 \; \left( \frac{0.01}{\eta_{Q}} \right)^{2/3} \left(\frac{0.1}{K}\right)^{1/2}  
\left(\frac{N_{\odot}}{10^{-12}}\right)^{2/3} \left(\frac{10.6}{\gamma_{frag}}\right)  \,{\rm km}  ~.\ee

The mass of the Q-ball is 
\be{as1a} M_{Q} = N_{\odot} M_{\odot} = 2.0 \times 10^{18} \;  \left(\frac{N_{\odot}}{10^{-12}} \right) \, {\rm kg}  ~.\ee  
Thus the mass of the $N_{\odot} = 10^{-12}$ Q-ball is similar to that of an asteroid of similar radius. 

The mass of the scalar, from \eq{c19a}, is 
\be{as2} m_{\phi} = 1.8 \times 10^{-20} \; \left( \frac{\eta_{Q}}{0.01} \right)^{2/3}
\left(\frac{10^{-12}}{N_{\odot}}\right)^{2/3} \left(\frac{10.6}{\gamma_{frag}}\right) \, \GeV  ~.\ee

The fragmentation temperature, from \eq{c26}, is 
\be{as3} T_{frag}  = 1.1 \times 10^{-3} \; \left( \frac{\eta_{Q}}{0.01} \right)^{2/3}
\left(\frac{10.6}{\gamma_{frag}}\right) 
\left(\frac{K}{0.1}\right)^{1/2}  
\left(\frac{10^{-12}}{N_{\odot}}\right)^{1/2} \, \GeV  ~.\ee

The Q-ball field, from \eq{c25}, is 
\be{as4} \phi_{0} = 3.7 \times 10^{11} \; \left( \frac{\eta_{Q}}{0.01} \right)^{1/3}
\left(\frac{10.6}{\gamma_{frag}}\right)^{1/2}  
\left(\frac{K}{0.1}\right)^{3/4}  
\left(\frac{10^{-12}}{N_{\odot}}\right)^{1/6} \, \GeV  ~.\ee
The initial field, from \eq{f5} and \eq{f6}, is then 
\be{as5}  \phi_{i} = 1.6 \times 10^{14} \; \left( \frac{\eta_{Q}}{0.01} \right)^{1/3}
\left(\frac{10.6}{\gamma_{frag}}\right)^{1/4}  
\left(\frac{10^{-12}}{N_{\odot}}\right)^{1/6} \, \GeV  ~.\ee

Since in this case the Q-balls are all of the dark matter, we need to check that the isocurvature perturbations are below the present observational limit \cite{planck}, 
\be{as6} \beta_{iso} \approx  \frac{P_{iso}}{P_{adi}} < 0.038  ~,\ee  
where $P_{iso}$ is the power spectrum of the Q-ball cold dark matter isocurvature perturbations and $P_{adi}$ is the power spectrum of the adiabatic perturbations, and we have assumed that $P_{iso} \ll P_{adi}$. Therefore we require that 
\be{as7} \frac{H_{I}}{\pi \phi_{i}}  < \beta_{iso}^{1/2} P_{adi}^{1/2} =  9.0 \times 10^{-6} ~,\ee
where we have set $P_{iso} =  4P_{\delta \phi}/\phi_{i}^{2}$ (since $\delta \rho_{\phi}/\rho_{\phi} = 2 \delta \phi/\phi$), with the power spectrum of the scalar field fluctuations is given by   $P_{\delta \phi} = H_{I}^{2}/4\pi^{2}$, and we have set $P_{adi}$ to its Planck value,  $2.1 \times 10^{-9}$ \cite{planck}. Therefore we require that the expansion rate during inflation satisfies   
\be{as8} H_{I} < 4.8 \times 10^{9} \; \left(\frac{\phi_{i}}{1.7 \times 10^{14} \GeV} \right) \, \GeV  ~.\ee 
Thus the energy scale of inflation has to be low, but this does not present any obstacle to the model. (Low values of $H_{I}$ are a feature of other ultra-light particle based models of dark matter, such as axions.) It will however restrict the tensor-to-scalar ratio to very small values, well below the values that will be observable by the next generation of CMB observations,
\be{as9} r \approx 3.8 \times 10^{-10} \; \left(\frac{H_{I}}{4.7 \times 10^{9} \GeV}\right)^{2} ~.\ee

This assumes that the collision rate of the Q-balls is less than the present expansion rate, so that Q-balls behave as collisionless cold dark matter. We will discuss the conditions for this to be true in the following subsection.

\subsection{Collision and Merger Rates} 

In order to check the consistency of the above possibilities, we need to check whether the Q-balls in the galaxy halo undergo collisions. We will assume that if Q-balls collide then they will merge. We will first consider the rate of Q-balls collisions assuming the local dark matter density and velocity. However, the density at the centre of galaxies may be much larger than the local density if an NFW profile, which behaves as $\rho_{DM}(r) \propto 1/r$ for small $r$, correctly describes the halo dark matter density. Thus it may be possible for Q-ball collisions and mergers to occur at the centre of the galaxy even if they do not occur at the local halo density. 

We assume that the Q-ball collision cross-section is the geometric cross-section $\sigma_{Q} = 4 \pi R_{Q}^{2}$. The collision rate is then 
\be{qc1} \Gamma_{Q} = n_{Q} \sigma_{Q} v  ~,\ee
where $v$ is the local dark matter velocity. The Q-ball number density is 
\be{qc2} n_{Q} = \frac{\Omega_{Q,\;0}\rho_{DM}}{\Omega_{DM,\, 0} M_{Q}}   ~,\ee
where $\Omega_{DM,\,0}$ is the present dark matter density parameter. 
We will use the typical values $\Omega_{DM,\,0} = 0.26$, $\rho_{DM} = 0.4 \GeV/{\rm cm^{3}} \equiv  3.07 \times 10^{-42} \GeV^4$ and $v = 250 {\rm km \, s^{-1}} \equiv 8.34 \times 10^{-4}$.

For the case of SMBH from Q-ball mergers, the mass $M_{Q}$ and radius $R_{Q}$ are given by \eq{sc11} and \eq{sc12}. Using these, we find that the Q-ball collision rate in the local halo is 
\be{qc3} \Gamma_{Q} = 1.8 \times 10^{-46}\,\epsilon_{merge}^{1/2} \left( \frac{\Omega_{\phi,\,0}}{2.6 \times 10^{-4}}\right)^{1/2}\left( \frac{0.1}{K}\right)^{3/4}  \left( \frac{0.01}{\eta_{Q}} \right) \left(\frac{\gamma_{frag}}{10.6}\right)^{3/2} \GeV ~.\ee 
The condition that collisions and mergers to have occurred is that $\Gamma_{Q} > H_{0}$, where $H_{0}$ is the present expansion rate, $H_{0} = 1,49 \times 10^{-42} \GeV$. This condition is satisfied if 
\be{qc4}   \epsilon_{merge}^{1/2} \left( \frac{\Omega_{Q,\,0}}{2.6 \times 10^{-4}}\right)^{1/2}\left( \frac{0.1}{K}\right)^{3/4}  \left( \frac{0.01}{\eta_{Q}} \right) \left(\frac{\gamma_{frag}}{10.6}\right)^{3/2} > 8.3 \times 10^{3}  ~.\ee
Since the LHS of this equation is unlikely to be significantly larger than 1, it is unlikely that mergers can occur in the local halo. However, it may be possible for mergers to occur at the centre of the galaxy if the dark matter density and so $n_{Q}$ is larger by a factor of $10^{4}$. 
  
For the case of asteroid mass Q-ball dark matter, the Q-ball radius and mass are given by \eq{as1} and \eq{as1a}. We assume that dark matter is entirely due to Q-balls, in which case $\Omega_{DM,\,0} = \Omega_{Q,\,0}$. The Q-ball collision rate is then 
\be{qc5} \Gamma_{Q} = 6.8 \times 10^{-49}  \left( \frac{0.01}{\eta_{Q}} \right)^{4/3} \left(\frac{0.1}{K}\right)  
\left(\frac{N_{\odot}}{10^{-12}}\right)^{1/3} \left(\frac{10.6}{\gamma_{frag}}\right)^{2} \GeV  ~.\ee
The condition that the Q-balls do not collide and so can behave as collisionless cold dark matter is that $\Gamma_{Q} < H_{0}$. This is satisfied if 
\be{qc6}   \left( \frac{0.01}{\eta_{Q}} \right)^{4/3} \left(\frac{0.1}{K}\right)  
\left(\frac{N_{\odot}}{10^{-12}}\right)^{1/3} \left(\frac{10.6}{\gamma_{frag}}\right)^{2}  < 2.2 \times 10^{6} ~.\ee 
Since the LHS of this equation is unlikely to be significantly larger than 1, it is likely that collisions of asteroid mass dark matter Q-balls do not occur in the local halo. 

The rate of collision of Q-balls in the case of SMBH from mergers is typically larger than that of the asteroid mass dark matter Q-balls, by a factor 260 (from the RHS of \eq{qc3} and \eq{qc5}). Thus there is a small window of densities at the centre of the galaxy for which Q-balls could merge to form SMBH black holes whilst asteroid mass black holes would not collide.  However, it is more likely that either SMBH via mergers or asteroid mass dark matter black holes is possible, but not both,  depending on the galaxy halo dark matter density profile.

\section{Conclusions} 

Our main objective here has been to determine if it is possible to set up a hidden sector scalar field that can consistently generate cosmologically significant Q-balls. The answer is yes, based on our analytical calculations of the linear growth of scalar field perturbations and of the properties of the Q-balls, and on inference from existing  numerical simulations of condensate fragmentation for the assumed scalar potential. We have focused on a scalar potential of a form that is well-known from flat directions in gravity-mediated supersymmetry breaking models, and for which analytical Q-ball solutions and numerical simulations of condensate fragmentation exist. Whilst the potential was originally motivated by supersymmetry, we note that the form of the potential can be equally well motivated by scale invariance that is broken by the mass squared of the scalar field. An important result of existing numerical simulations is that even if the initial condensate is neutral, the final state after fragmentation will consist of Q-balls, as the condensate will initially fragment to neutral oscillons that break up into a distribution of positive and negative Q-ball pairs. This means that the scalar potential can be quite simple, with no need for any initial Affleck-Dine generation of a charge asymmetry.

The Q-balls typically form during the radiation-dominated era and therefore exist throughout the growth of density perturbations and the formation of galaxies. They could have a range of applications, from very large and heavy Q-balls (O(100) light year radius and $10^{6} M_{\odot}$ mass), with around one per galaxy, that might enhance SMBH and galaxy formation, to smaller Q-balls (of roughly lunar mass but solar radius) that could merge within a galaxy to form a $10^{6} M_{\odot}$ SMBH. Whether this can happen in practice requires a detailed numerical study of the evolution of structure when superheavy Q-balls contribute to the dark matter density. Finally, it is possible for Q-balls of asteroid size and mass to account for all of the dark matter. In this case isocurvature perturbations of the Q-ball density require that the expansion rate during inflation is less than around $10^{10} \GeV$, implying a very small tensor-to-scalar ratio, $r < 10^{-9}$.   

This is very much a first study of hidden sectors that can produce cosmologically significant Q-balls. The potential we have considered is quite simple and it would be interesting to revisit the simulations of condensate fragmentation for this potential, as the most recent existing simulation is now more than a decade old. It also would be interesting to see if other scalar potentials can lead to interesting oscillons or Q-balls in the present Universe. In addition, it would be interesting to investigate whether there is a gravitational wave signature of the condensate fragmentation process. Finally, it is necessary to determine via a numerical simulation how a cosmological density of such objects evolves during structure formation and whether they can enhance SMBH and galaxy formation as compared to conventional cold dark matter.

   \renewcommand{\theequation}{A-\arabic{equation}}
 \setcounter{equation}{0}  

\section*{Appendix A: Dominance of self-interaction over gravity} 

Our analysis assumes that the Q-ball is dominated by the attractive self-interaction between the scalars and that gravity can be neglected when computing Q-ball properties. To check this, we will compare the binding energy of the Q-ball due to the self-interaction and the Newtonian gravitational potential energy of the Q-ball.  

The effective mass of the scalars is 
\be{ap1} m_{\Phi,\,eff}^{2} = m_{\Phi}^{2}\left(1 - K \ln\left(\frac{\phi^{2}}{\mu^{2}}\right) \right)  ~.\ee
In the vacuum, we have $\phi \rightarrow 0$ and so $m_{\Phi,\,eff} \rightarrow \infty$. However, we expect the $\phi$ dependence of the logarithmic term to change as $\phi \rightarrow 0$. In the case of MSSM flat directions, the potential is run to small $\phi$ using the renormalisation group equations, and eventually the bare mass terms of the particles contributing to the Coleman-Weinberg potential dominates their $\phi$ dependent mass, cutting off the $\phi$ dependence of the logarithm. Here we will simply assume that the logarithm cuts off at a value $-A$, where $A > 0$. Then as $\phi \rightarrow 0$, the effective mass of $\phi$ scalars in the vacuum becomes  
\be{ap2} m_{\Phi,\,vac}^{2} = \left(1 + KA \right)m_{\Phi}^{2}  ~.\ee
The Q-ball binding energy due to the self-interaction is then 
\be{ap3} E_{B} = m_{\Phi,\,vac}Q - M_{Q} = \left[ \left(1 + AK\right)^{1/2} \left(1 + 2K\right)^{1/2} - \left(1 + \frac{5 K}{2}\right) \right] \frac{\pi^{3/2} \phi_{0}^{2}}{K^{3/2} m_{\phi}}   ~,\ee 
where $Q$ and $M_{Q}$ are given by \eq{e9} and \eq{e10}. 
If $K$ and $AK$ are small compared to 1 then this becomes 
\be{ap4} E_{B} \approx \frac{(A - 3) \pi^{3/2} \phi_{0}^{2}}{2 K^{1/2} m_{\phi}} ~,\ee
where $A > 3$ for the Q-ball to be stable. The binding energy per charge is then (where $Q$ is given by \eq{e9})   
\be{ap4a} \frac{E_{B}}{Q} = \frac{\left(A-3\right) K m_{\phi}}{2 \left(1 + 2K\right)^{1/2} } ~.\ee
So $E_{B}/Q \sim K m_{\phi}$ when $K$ is small compared to 1.  The gravitational potential energy of the Q-ball is 
\be{ap5} V = \int_{0}^{\infty} 4 \pi r G M_{Q}(r) \rho_{Q}(r) dr   ~,\ee
where $M_{Q}(r)$ is the mass of the Q-ball within radius $r$, 
\be{ap5a} M_{Q}(r) = \int_{0}^{r} 4 \pi r^{2} \rho_{Q}(r) dr ~.\ee 
The energy density of the Q-ball, $\rho_{Q}(r)$, can be read off from the integral in \eq{e10}
\be{ap5b} \rho_{Q}(r) =  \phi_{0}^{2} \left( \frac{m_{\phi}^{2} + \omega^{2}}{2} + K^{2} m_{\phi}^{4} r^{2} \right) e^{-K m_{\phi}^{2} r^{2}}   ~.\ee 
We will overestimate the gravitational potential energy in order to derive a conservative condition for the Q-ball to be dominated by the scalar self-interaction. The Q-ball energy density rapidly goes to zero once $r^{2} > (Km_{\phi}^{2})^{-1}$ due to the exponential factor, so we can set the upper limit of the integral to the Q-ball radius, $R_{Q} = \sqrt{2}/(\sqrt{K} m_{\phi})$ and overestimate the integral by setting the exponential factor to 1.
In addition, the largest value of the $K^{2} m_{\phi}^{4} r^{2}$ term is $K^{2} m_{\phi}^{4} R_{Q}^{2} = 2 K m_{\phi}^{2}$. So we can overestimate the gravitational potential by setting the $K m_{\phi}^{4} r^{2}$ term to its largest value. So for $r \leq R_{Q}$ we use  
\be{ap6} \rho_{Q}(r) =  \phi_{0}^{2} \left( \frac{m_{\phi}^{2}(1 + 4 K) + \omega^{2}}{2}\right)   ~,\ee 
a constant, and set $\rho_{Q} = 0$ for $r > R_{Q}$. 
Then
\be{ap7} M_{Q}(r) =  \frac{4 \pi \rho_{Q} r^{3}}{3} ~\ee  
for $r \leq R_{Q}$, and  
\be{ap8} V = \int_{0}^{R_{Q}} 4 \pi r G M_{Q}(r) \rho_{Q} dr = \frac{16 \pi^{2} G \rho_{Q}^{2} R_{Q}^{5}}{15} =  \frac{2 \sqrt{2} \pi \phi_{0}^{4}}{15 K^{5/2} M_{Pl}^{2}} \frac{\left(m_{\phi}^{2}(1 + 4 K) + \omega^{2}\right)^{2}}{m_{\phi}^{5}}  ~.\ee   
For small $K$, so that $\omega^{2} \approx m_{\phi}^{2}$,  this becomes 
\be{ap9} V \approx  \frac{8 \sqrt{2} \pi \phi_{0}^{4}}{15 K^{5/2} M_{Pl}^{2} m_{\phi}}  ~.\ee   
The condition that the (overestimated) gravitational potential energy of the Q-ball is less than its binding energy, $V < E_{B}$,  is then 
\be{ap10} \frac{\phi_{0}}{M_{Pl}} < \left(\frac{15 \sqrt{\pi} (A - 3)}{16 \sqrt{2}}\right)^{1/2} K \approx 1.1 (A-3)^{1/2} K   ~.\ee 
Therefore as long as $\phi_{0}/M_{Pl} < K$, the self-interaction will dominate gravity in the Q-ball.

\end{document}